# Vindication of Quantum Locality[1]

David Deutsch

Centre for Quantum Computation
The Clarendon Laboratory
University of Oxford, Oxford OX1 3PU, UK

July 2011, Revised September 2011

*In a previous paper Hayden and I proved, using the Heisenberg picture, that quantum physics satisfies Einstein's criterion of locality. Wallace and Timpson have argued that certain transformations of the Heisenberg-picture description of a quantum system must be regarded as leaving invariant the factual situation being described, and that taking this into account reveals that Einstein's criterion is violated after all. Here I vindicate the proof and explain some misconceptions that have led to this and other criticisms of it.*

## 1. The proof

Einstein's (1949) criterion for locality is that for any two spatially separated physical systems $\mathfrak{S}_1$ and $\mathfrak{S}_2$, 'the real factual situation of the system $\mathfrak{S}_2$ is independent of what is done with the system $\mathfrak{S}_1$'. A previous paper (Deutsch & Hayden (2000)) included a proof that quantum physics satisfies this criterion. The method was first to prove that every quantum computational network satisfies it, and then to infer the same for general quantum systems by appealing to the universality of such networks[2]. For convenience I summarise the proof here:

Consider a quantum computational network $\mathfrak{A}$ containing $n$ qubits $\mathbb{Q}_1, \ldots, \mathbb{Q}_n$. In the Heisenberg picture, each qubit $\mathbb{Q}_a$ at time $t$ can be described by a triple

---

[1] To be published by Proc. R. Soc. A.

[2] Despite the discreteness of the observables of qubits, this universality includes the ability to simulate systems with nominally continuous degrees of freedom. That is because the existence of the Bekenstein bound (1981) implies that the state space of any quantum system of finite volume has finite dimension. For a quantum-field-theoretic version of our proof that does not rely on the Bekenstein bound, see Rubin (2002, 2011).

$$\hat{\mathbf{q}}_a(t) = \left(\hat{q}_{ax}(t), \hat{q}_{ay}(t), \hat{q}_{az}(t)\right) \tag{1}$$

of its observables, satisfying

$$\left.\begin{array}{l} \left[\hat{\mathbf{q}}_a(t), \hat{\mathbf{q}}_b(t)\right] = 0 \quad (a \neq b), \\ \hat{q}_{ax}(t)\hat{q}_{ay}(t) = i\hat{q}_{az}(t) \\ \hat{q}_{ax}(t)^2 = \hat{1} \end{array}\right\} \text{(and cyclic permutations over } x, y, z\text{).} \tag{2}$$

Choose, as the computation basis at time $t$, the simultaneous eigenstates $|z_1, \ldots, z_n; t\rangle$ of the observables

$$\hat{z}_a(t) = \tfrac{1}{2}\left(\hat{1} - \hat{q}_{az}(t)\right). \tag{3}$$

So $\hat{z}_a(t)$ is the projector for the $a$'th qubit to hold the value 1 at time $t$.

The effect of a gate **G** acting on $k$ qubits between times $t$ and $t+1$ is

$$\hat{\mathbf{q}}_{a^G}(t+1) = \mathsf{U}_G^\dagger\left(\hat{\mathbf{q}}_{1^G}(t), \ldots, \hat{\mathbf{q}}_{k^G}(t)\right) \hat{\mathbf{q}}_{a^G}(t) \mathsf{U}_G\left(\hat{\mathbf{q}}_{1^G}(t), \ldots, \hat{\mathbf{q}}_{k^G}(t)\right), \tag{4}$$

where $1^G, \ldots k^G$ are the indices of the qubits that are acted upon by **G**, and $a^G$ is any such index. All observables of other qubits are unaffected by **G**. The form of each $\mathsf{U}_G$ qua function of its arguments is fixed and characteristic of the corresponding gate **G**. Its form qua unitary transformation varies accordingly. Conversely, given the form of all the $\hat{\mathbf{q}}_a(t+1)$ as functions of the $\hat{\mathbf{q}}_a(t)$, one can infer what gates are present during that computational step and which qubits pass through which gate. In other words, the dynamics – the laws of motion – of the qubits of the network are encoded in the histories of their Heisenberg observables.

For present purposes we lose no generality by assuming that the Heisenberg state of the qubits of $\mathfrak{A}$ is $|0\rangle$, the zero-eigenvalue eigenstate of all the $\hat{z}_a(0)$. For if we wanted to study an arbitrary network $\mathfrak{A}'$ in some other Heisenberg state, say

$|\Psi\rangle$, we could do so by prepending to $\mathfrak{N}'$ a network $\mathfrak{N}''$ that has the same number of qubits and effects a unitary transformation mapping $|0\rangle$ to $|\Psi\rangle$ (such a network must exist because of universality) and is in its Heisenberg state $|0\rangle$. That means that the combined network $\mathfrak{N}$ would also be in the state $|0\rangle$, and $\mathfrak{N}''$ would prepare $\mathfrak{N}'$ in the Schrödinger state $|\Psi\rangle$ at the time when $\mathfrak{N}'$ begins to compute, which is the same as preparing it in its Heisenberg state $|\Psi\rangle$. A proof that the combined network in state $|0\rangle$ satisfies Einstein's criterion is then also a proof that $\mathfrak{N}'$ in state $|\Psi\rangle$ does. Consequently we can confine attention to networks in state $|0\rangle$, and we can adopt the compact notation $\langle \hat{A}(t) \rangle \equiv \langle 0|\hat{A}(t)|0\rangle$ for the expectation values of observables.

Rather than engage with the complication of modelling the relevant properties of spacetime, such as the finiteness of the speed of light, we prove a stronger result than strictly required, replacing 'spatially separated' in Einstein's criterion by 'mutually isolated'. Qubits $\mathbb{Q}_a$ and $\mathbb{Q}_b$ are mutually isolated during a particular computational step starting at time $t_1$ if (though not necessarily only if) they are not passing through the same gate during that step. $\mathbb{Q}_b$ then remains isolated from $\mathbb{Q}_a$ at least until it enters a gate with any qubit which, at some time after $t_1$, was not isolated from $\mathbb{Q}_a$.

'Anything done with' the qubits of a sub-network $\mathfrak{N}_1$ means any locally-caused effect on them. We build that into our model by considering networks $\mathfrak{N}(\theta)$ containing one or more parameterised gates in a sub-network $\mathfrak{N}_1(\theta)$. The parameters symbolised by $\theta$ are deemed to be set by an external agent, such as an experimenter trying to cause a violation of Einstein's criterion. Any preparation supposedly needed to cause a later violation is represented by a sub-network $\mathfrak{N}''$ acting before $\mathfrak{N}_1(\theta)$ begins. (Modelling a system in a state other than $|0\rangle$ is formally the same as modelling such a preparation.) Einstein's criterion then becomes: Let $\mathfrak{N}_1(\theta)$ and another sub-network $\mathfrak{N}_2$ of $\mathfrak{N}(\theta)$ be mutually isolated during at least the period when the gates parameterised by θ act. Then, for as long as they remain mutually isolated, the factual situation of the qubits of $\mathfrak{A}_2$ cannot depend on θ.

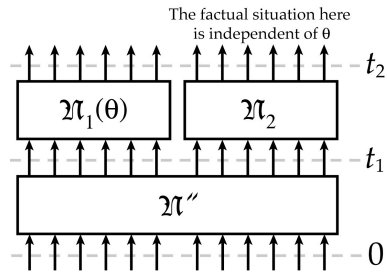

Fig. 1: Einstein's criterion

Since the state $|0\rangle$ and the prepended network $\mathfrak{A}''$ contain no information about θ, and since the dynamics of each gate are encoded in the change in the observables of the qubits on which that gate acts, the full factual situation of any sub-network of the network is specified by the observables of that sub-network alone. And since the observables change only inside gates, Einstein's criterion is necessarily satisfied by every quantum computational network, and hence by all quantum systems.

## 2. The density operator

In classical physics, the complete description of a physical system can be separated into a kinematical part, which consists at each instant of a set of real numbers that are in principle measurable (such as particle positions or field values), and a dynamical part, which consists of a rule (such as a differential equation) stating how the kinematical part changes with time. An important departure from this in quantum physics is that observations do not generally have quantum observables as outcomes, despite the name. Instead, the connection between theory and experiment is via the observables' expectation values – itself a somewhat misleading term since it suggests that observables are stochastic variables (see Section 9). Similarly, there is a potential ambiguity when we speak of 'measuring the value' of an observable, for there are two quite different things that the term 'value' might refer to. One is the observable's Hermitian-operator value. The other is a real number – an outcome of a measurement of the observable.

In the Heisenberg picture, observables are functions of time just as in classical physics; but unlike in classical physics they specify both kinematics and dynamics. Expectation values are given by a fixed linear function, $\langle \cdot \rangle$, from observables to real numbers. In the Schrödinger picture, kinematics and dynamics are separate, as in classical physics, but unlike in classical physics the kinematical part does not consist of the observables (which never change), but of density operators, which specify the changing expectation values of observables. Every quantum system or subsystem $\mathfrak{S}$ has a density operator $\hat{\rho}_{\mathfrak{S}}(t)$, which has the property $\text{Tr}\,\hat{\rho}_{\mathfrak{S}}(t)\hat{A} = \langle \hat{A}(t) \rangle$ for any observable $\hat{A}(t)$ of $\mathfrak{S}$, where $\hat{A} \equiv \hat{A}(0)$ is the Schrödinger-picture form of the observable. For example, the density operator of $\mathfrak{Q}_a$ at time $t$ is $\hat{\rho}_a(t) = \frac{1}{\text{Tr}\hat{1}}\left(\hat{1} + \sum_i \hat{\mathbf{q}}_{ai}(0)\langle \hat{\mathbf{q}}_{ai}(t)\rangle\right)$.

## 3. Wallace and Timpson's critique

Given a network $\mathfrak{A}$ in state $|\mathbf{0}\rangle$, consider any unitary-operator-valued function of time $\mathsf{V}(t)$ such that for some real function $\phi(t)$,

$$\mathsf{V}(t)|\mathbf{0}\rangle = e^{i\phi(t)}|\mathbf{0}\rangle. \tag{5}$$

Then define

$$\hat{q}'_{ai}(t) = \mathsf{V}^{\dagger}(t)\hat{q}_{ai}(t)\mathsf{V}(t). \tag{6}$$

Since every observable $\hat{A}(t)$ of $\mathfrak{A}$ is some algebraic function of the $\hat{q}_{ai}(t)$, (5) makes all expectation values $\langle \hat{A}(t) \rangle$ invariant under the replacements $\hat{\mathbf{q}}_a(t) \to \hat{\mathbf{q}}'_a(t)$. Hence Wallace and Timpson claim that 'nothing whatever – no observable data, no theoretical considerations – can tell us that the physical state is given by [the $\hat{\mathbf{q}}_a(t)$] rather than [the $\hat{\mathbf{q}}'_a(t)$]'. They draw an analogy with gauge transformations in classical electrodynamics, where no experiment can detect whether the vector potential is $A_\mu(x)$ or $A'_\mu(x) = A_\mu(x) + \nabla_\mu \varphi(x)$, where $x$ represents position in spacetime and $\varphi(x)$ is an arbitrary scalar field. Just as we regard all vector potentials related by such transformations as describing physically the same electromagnetic field, so, they argue, we must regard any two Heisenberg-picture descriptions that are related by a Wallace–Timpson transformation as referring to the same factual situation.

It would then follow that the full factual situation of a quantum system $\mathfrak{S}$ is represented by the density operator of the combined system of $\mathfrak{S}$ and everything it is entangled with. Only part of that information, namely the density operator of $\mathfrak{S}$ itself, is 'independent of what is done' to other systems isolated from $\mathfrak{S}$. The rest, which exists if and only if $\mathfrak{S}$ is entangled with another system $\mathfrak{S}'$, can depend on what is done to $\mathfrak{S}'$ even when it is spatially separated from $\mathfrak{S}$. That is the origin of the historical claim that entangled quantum systems violate Einstein's criterion. And so Wallace and Timpson conclude that 'we are again left with a theory whose states are non-local'.

4. **The meaning of Wallace–Timpson transformations**

What, in the Heisenberg-picture description, does the Wallace–Timpson identification rule deny is part of the factual situation of a quantum system? Consider a two-qubit network whose density operator $\hat{\rho}(t)$ evolves as follows, expressed in terms of eigenstates of the $\hat{z}_a(0)$:

$$\hat{\rho}(0) = |0,0\rangle\langle 0,0|; \quad \hat{\rho}(1) = |1,0\rangle\langle 1,0|; \quad \hat{\rho}(2) = |1,1\rangle\langle 1,1|. \tag{7}$$

Since the states in (7) are pure, it must, according to the Wallace–Timpson identification, be a complete description of the factual situation of both qubits. That would mean that the infinity of *networks* that could cause this evolution

would all give rise to the same factual situation of the *qubits* over time. Here are two such networks:

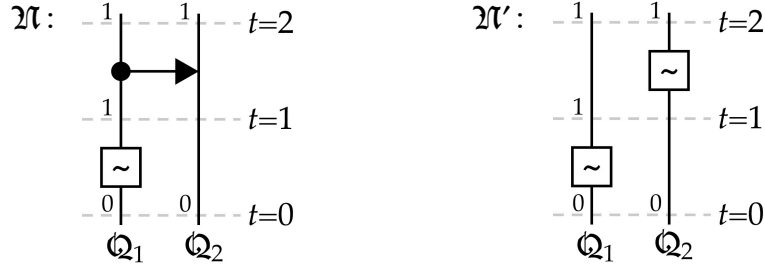

Fig. 2: Two networks related by a Wallace–Timpson transformation

(The horizontal arrow represents a *controlled-not* gate and '~' labels a *not* gate.) Since a *not* gate effects the unitary transformation $\hat{q}_{ax}(t)$ on $\mathbb{Q}_a$ and a *controlled-not* gate (with $\mathbb{Q}_a$ as the control and $\mathbb{Q}_b$ as the target) effects a different transformation $\tfrac{1}{2}\left(\hat{\mathbf{1}} + \hat{q}_{bx}(t) + \hat{q}_{az}(t) - \hat{q}_{az}(t)\,\hat{q}_{bx}(t)\right)$, the Heisenberg-picture descriptions of the qubits in the two networks are different:

$$\mathfrak{N}:\begin{cases}\hat{\mathbf{q}}_1(2) = \left(\hat{q}_{1x}\hat{q}_{2x},\, -\hat{q}_{1y}\hat{q}_{2x},\, -\hat{q}_{1z}\right)\\ \hat{\mathbf{q}}_2(2) = \left(\hat{q}_{2x},\, -\hat{q}_{1z}\hat{q}_{2y},\, -\hat{q}_{1z}\hat{q}_{2z}\right)\end{cases} \qquad \mathfrak{N}':\begin{cases}\hat{\mathbf{q}}'_1(2) = \left(\hat{q}'_{1x},\, -\hat{q}'_{1y},\, -\hat{q}'_{1z}\right)\\ \hat{\mathbf{q}}'_2(2) = \left(\hat{q}'_{2x},\, -\hat{q}'_{2y},\, -\hat{q}'_{2z}\right)\end{cases} \qquad (8)$$

where all the observables on the right-hand sides of the equations are evaluated at $t=0$ and the parameters '(0)' have been suppressed for brevity. A Wallace–Timpson transformation relating those two descriptions is

$$\mathsf{V}(0) = \hat{\mathbf{1}};\quad \mathsf{V}(1) = \hat{\mathbf{1}};\quad \mathsf{V}(2) = \tfrac{1}{2}\left(\hat{\mathbf{1}} + \hat{q}_{1z} + \hat{q}_{2x} - \hat{q}_{1z}\hat{q}_{2x}\right). \qquad (9)$$

Thus, the elements of reality whose existence the Wallace–Timpson identification rule denies are the different *dynamics* of quantum systems – their different laws of motion, or in the network model, their different gates. Let me note in passing what a bizarre world-view would be forced on us by this rule. For instance, it would imply that a sufficiently detailed three-dimensional recording of a person, when played, would *be* a person: a Wallace–Timpson transformation, just like (9) only more complicated, relates the two, so the rule calls them factually identical situations. More generally, banishing laws of motion from the realm of physical reality means declaring that reality consists only of *what* happens and not *why*, which in turn means banishing almost all explanation from science. But in the event, we shall see that (7) is an incomplete representation of *what* happens, so this philosophical issue does not arise.

Returning now to the issue of Einstein's criterion, consider: are $\mathbb{Q}_1$ and $\mathbb{Q}_2$ in Fig. 2 mutually isolated during the period $1 < t < 2$? In network $\mathfrak{N}$ they are not, since $\mathbb{Q}_2$, the target of the *controlled-not* operation, changes its stored value from 0 to 1 only because $\mathbb{Q}_1$ holds a 1 during that period. But in network $\mathfrak{N}'$ they are isolated, since each is acted upon by its own single-qubit gate independently of the other. Similarly, does the factual situation of $\mathbb{Q}_2$ at $t = 2$ 'depend on what was done' to $\mathbb{Q}_1$ during the period $0 < t < 1$ (namely that its value was flipped from 0 to 1)? Again, the answer is different for $\mathfrak{N}$ and $\mathfrak{N}'$. Therefore, according to the Wallace–Timpson identification, there is no factual difference between these two qubits being isolated or interacting, nor between one of them being affected or unaffected by what is done to the other – nor, therefore, between satisfying and violating Einstein's criterion.

Nor is that just a peculiarity of those particular networks. For any network $\mathfrak{N}$, consider any two descriptions related by a Wallace–Timpson transformation. The laws of motion for the $\hat{\mathbf{q}}'_a(t)$ are different from those for the $\hat{\mathbf{q}}_a(t)$. Instead of (4), they are, for each $k$-qubit gate $\mathbf{G}$ of $\mathfrak{N}$,

$$\hat{\mathbf{q}}'_{a_G}(t+1) = \mathsf{V}^\dagger(t+1)\mathsf{V}(t)\mathsf{U}^\dagger_\mathbf{G}\left(\hat{\mathbf{q}}'_{1_G}(t),\ldots,\hat{\mathbf{q}}'_{k_G}(t)\right)\hat{\mathbf{q}}'_{a_G}(t)\mathsf{U}_\mathbf{G}\left(\hat{\mathbf{q}}'_{1_G}(t),\ldots,\hat{\mathbf{q}}'_{k_G}(t)\right)\mathsf{V}^\dagger(t)\mathsf{V}(t+1). \quad (10)$$

For generic choices of $\mathsf{V}(t)$, (10) does not describe a $k$-qubit gate. It acts non-trivially on all $n$ qubits of $\mathfrak{N}'$ because the factor $\mathsf{V}^\dagger(t+1)\mathsf{V}(t)$, expressed in terms of the primed observables $\hat{q}'_{ai}(t)$, generically depends on all $3n$ of them. The same holds for all the other gates acting during the same period, and therefore, since a particular qubit can be in only one gate during any computational step, the primed observables can describe a network with not only different gates but different topologies (how the gates are connected and how many qubits each acts on) and hence different locality structures (which qubits are mutually isolated). Generically they describe networks in which only a single gate acts at a time, on all *n* primed qubits simultaneously, so that even if $\mathfrak{A}$ models a system consisting of several mutually isolated subsystems, $\mathfrak{A}'$ models one in which every qubit affects every other during every computational step.

Hence, the Wallace–Timpson identification rule would compel us to regard the distinctions drawn by Einstein's criterion as not referring to anything real. That, and not a violation of the criterion as such, is the content of Wallace and Timpson's conclusion that 'quantum physics has non-local states'.

Nevertheless, what about their argument that their identification *must* be made, because the two descriptions result in identical predictions, just as two different gauges do in electrodynamics?

5. **Experimental tests in quantum physics**

A statement, such as $\langle \hat{A}(t) \rangle = k$, assigning a value to a symbol in a scientific theory, constitutes a prediction only if the theory (or some associated theory) also describes a physical procedure for testing that prediction. That description is the *operational meaning* of the prediction.

Let me stress immediately that I am not claiming that the operational meaning is the *only* content of a scientific theory (or equivalently that two theories necessarily describe the same reality if they make identical predictions). That would be positivism, which is a catastrophic philosophical error (see Deutsch 2011). But that philosophical issue need not detain us here, because I shall now show that it is not the case that two theories related by a Wallace–Timpson transformation make identical predictions.

At a given time $t_1$ the density operator $\hat{\rho}_\mathfrak{S}(t_1)$ summarises all predictions of expectation values $\langle \hat{A}(t_1) \rangle$ of observables $\hat{A}(t_1)$ of $\mathfrak{S}$ at time $t_1$. The procedure for testing a prediction $\langle \hat{A}(t_1) \rangle = k$ includes measuring $\hat{A}(t_1)$ instantaneously at time $t_1$ on many identically prepared instances of $\mathfrak{S}$, using instruments not initially entangled with any of those instances. Here 'instantaneously' means rapidly compared with the timescale on which $\mathfrak{S}$ would change in the absence of a measurement. Then one calculates whether the mean value of the outcomes differs significantly from *k*.

However, after one has performed a measurement on one of those instances of $\mathfrak{S}$, one can wait until a later time $t_2$ and then perform another. The expectation value of the second outcome cannot in general be calculated from the originally-calculated $\hat{\rho}_\mathfrak{S}(t_2)$ alone, because the first measurement process is itself an interaction that changes the density operator from what it would have been in the absence of that measurement.

In the network model, we can represent this broader class of experiments by including extra qubits (to hold the outcomes of measurements) and interposing extra gates between the gates of a network (to represent the measurement interactions). All such measurements are therefore already included in the general scheme described in Section 1, of considering parameterised sub-networks $\mathfrak{N}_1(\theta)$. By including, in the set of networks accessible by varying θ, arbitrary measurement gates before and after any particular gate (for example at times 1 and 2 in each network in Fig. 2), with ancillary qubits to store the outcomes of those measurements, and by running $\mathfrak{N}(\theta)$ repeatedly for various values of θ, one can model experiments that determine the dynamics of the gate to any desired accuracy. Hence they also determine the operator values of all the observables in terms of what they were at $t = 0$ (in other words, the complete operator algebra can be determined), and in this way all the information in the Heisenberg-picture representation can be tested by experiment.

So it is not true that 'nothing whatever – no observable data, no theoretical considerations – can tell us' whether the factual situation is given by the original theory or the transformed one. Those two cases can be straightforwardly distinguished by the very equipment needed to measure the expectation values – which the Wallace–Timpson identification says represent the only elements of reality. This is not the case with the classical electromagnetic vector potential: no classical equipment could measure it. So whether or not it makes sense to regard it as representing an element of reality[3], in quantum theory there is no option but to do so for the Heisenberg observables.

Logically, that completes the vindication of the Deutsch–Hayden proof. But there are some explanatory loose ends, which I shall address in the remaining sections.

## 6. A new quantum-theoretic picture?

Nothing in the above argument depends on adopting the Heisenberg picture. It is convenient to do so, since it displays the flow of information in quantum systems so explicitly; but the fact that the Heisenberg observables contain information that is not contained in any of the density operators, but does describe the reality of a quantum system, can be deduced in the Schrödinger picture as well. All the above arguments can be straightforwardly translated into the Schrödinger picture: one begins by defining operator-valued variables equal to the Heisenberg-picture observables of each qubit; then one proves that in the networks used in the proof, those variables can be affected only by local interactions of that qubit, but contain all the information in the Schrödinger state; then one proceeds as in Section 1.

---

[3] Whether the quantum-mechanical Aharonov-Bohm effect forces us to regard the vector potential as an element of reality too is beyond the scope of this paper.

Some authors (e.g. Vedral & Horsman 2006) have described our method of using a standard Heisenberg state $|0\rangle$ as a new quantum-theoretic picture, or at least as a variant of the Heisenberg picture. Timpson (2005) even considers it a new theory purporting to replace quantum theory. But in fact we are using a standard Heisenberg picture and merely *confining attention* to networks that are in state $|0\rangle$. Such networks suffice to prove our result because, as I pointed out in Section 1, any network in an arbitrary Heisenberg state $|\Psi\rangle$ can be regarded as a sub-network of one in state $|0\rangle$, and if any network satisfies Einstein's criterion then so does every sub-network of it. Hence we are neither making any assumption nor losing any generality. We are working in a standard Heisenberg picture, equivalent to the Schrödinger picture in every way except for the inconvenience of the latter for analysing information flow.

## 7. Wallace–Timpson transformations and operational meanings

Under a transformation of variables in a physical theory, operational meanings do not in general transform trivially. For example, if $x_1(t),\ldots,x_n(t)$ are the positions of $n$ Newtonian billiard balls, then the operational meaning of $x_1(t) = k$ might include something like: 'put a detector at the position $k$ at time $t$, and observe whether it fires'. But if we then define

$$x'_i(t) = \sum_j \mathsf{V}^j_i(t) x_j(t) \tag{11}$$

for some time-dependent invertible matrix $\mathsf{V}^j_i(t)$, there is no way to test whether $x'_1(t) = k$ merely by putting a measuring instrument at a particular place. It involves something like quickly searching the whole of the available space for the balls – or perhaps searching for $n-1$ of them and then placing a detector at the predicted position of the remaining, $m$'th ball: $\left(k - \sum_{i \neq m} \mathsf{V}^i_m(t) x_i(t)\right)/\mathsf{V}^m_m(t)$.

If we transform all the operational meanings in that way when we transform the variables, the transformed theory describes the same factual situation as the original. In particular, if it happened to be more convenient in some situations to use the primed variables (for instance, the position of the centre of mass is such a variable), then one could use the primed theory to prove that the unprimed variables describe localised objects and then that the system satisfies Einstein's criterion, just as one can use the Schrödinger picture to prove that quantum theory does. So I shall call that the *content-preserving* transformation. It is not what Wallace and Timpson use. They use what I shall call the *formal* transformation of operational meanings: simply taking the operational meanings from the original theory and putting a prime on each unprimed observable. So the resulting theory says that the $x'_i(t)$, and not necessarily the $x_i(t)$, are the positions of billiard balls. That does not, in general, describe the same factual situation.

But, say Wallace and Timpson, sometimes it does – specifically, when the transformed theory predicts the same values for all the transformed variables as the original theory did for the original variables. So, to complete the analogy between (11) and the Wallace–Timpson transformation (6), let us choose $\mathsf{V}^j_i(t)$ such that the $x_i(t)$ constitute a unit-eigenvalue eigenvector of it. Then $x'_i(t) = x_i(t)$ for all *t*, and so all three theories – the original one and the transformed ones under both the content-preserving and the formal operational meanings – say that there is a particle at each $x_i(t)$.

Yet the formally-transformed theory still describes a different physical system from the other two. Although its 'billiard balls' are moving on the same trajectories as the original ones, they are obeying different laws of motion. For example, suppose that the balls in the original theory move in straight lines unless $|x_i(t) - x_j(t)| = 2r$, in which case they bounce specularly. Then in the formally transformed theory the primed balls numbered *i* and *j* suddenly change course whenever $\left|\sum_k \left(\mathsf{V}^{-1\,k}_i(t) - \mathsf{V}^{-1\,k}_j(t)\right) x'_k(t)\right| = 2r$, a condition that generically depends on the positions of *all* the (primed) balls and *could* be satisfied even if they were

nowhere near each other. So their law of motion is, like (10), non-local. As with (10), and with the pair of networks in Fig. 2 in particular, that drastic difference between the formally transformed theory and the original one is not apparent if the balls are set in motion with the specified initial conditions and then left alone, because for that situation, the two different laws of motion happen to produce the same trajectories. But if we were free to experiment on both systems, and were to perturb their corresponding variables $x'_i(t)$ by the same amounts $\Delta x'_i(t)$ such that they no longer constituted an eigenvector of $V^j_i(t)$, their respective trajectories would in general differ subsequently.

Given *any* extended physical system that satisfies Einstein's criterion, there exist descriptions of it in which the equations of motion for some variables depend on all the variables. So that property is not a reasonable criterion for non-locality. Moreover, it is a property of formalisms but is not meaningful for physical systems, while Einstein's criterion is about physical systems only.

An important class of operational meanings are the specifications of which physical processes constitute measurements, and of what they measure. Everett (1957) was among the first to take seriously that measurements are subject to the same laws of physics as all other processes and have to be analysed as such in order for the resulting theory to be consistent. As Wallace and Timpson rightly remark: "in a physical theory we would normally prefer that what is 'observable' (i.e. the expectation values derived from $|0\rangle$) would emerge from a physical analysis of measurement, rather than by *fiat*". But they themselves do not implement this. Doing so would, among other things, entail recognising that under their formal transformation of operational meanings, measurement interactions do not map to measurement interactions. The networks of Fig. 2, related by the Wallace–Timpson transformation (9), are a simple instance of this: the last step of $\mathfrak{N}$ constitutes a perfect measurement of $\hat{z}_1(1)$, while $\mathfrak{N}'$ makes no measurement at all. This would make the Wallace–Timpson identification impossible even if nothing else did, because two theories that disagree about what constitutes a measurement of what, cannot be said to 'make identical predictions'.

## 8. The counterfactual import of Einstein's criterion

Einstein's criterion requires the factual situation of $\mathcal{S}_2$ not to depend on what is done to $\mathcal{S}_1$. Thus it refers to variants of $\mathcal{S}_1$ – systems to which 'something had been done' that would make them different from $\mathcal{S}_1$ in its actual condition – and to what $\mathcal{S}_2$ would or would not do in that counterfactual situation.

We have seen that the Wallace–Timpson identification erases that distinction between dependent and independent, and between mutually isolated and interacting, and hence between satisfying Einstein's criterion or not (erasing the latter distinction being the purpose of the identification). Also between local and non-local *dynamics*, and between different dynamical laws that happen to produce the same motion in a given instance of a physical system, and between what does or does not constitute a measurement of anything, and even between people and recordings. All these erased distinctions involve that same counterfactual element. And we can add *information* and *causation* to the list.

In the network model, we incorporated such comparisons by means of the externally set parameters θ. However, the external agent is itself a physical system. If the rebuttal of Wallace and Timpson's argument depends on analysing how a system responds to the actions of an external agent, can it be evaded by applying their argument to the combined system including the agent, and ultimately to the entire physical world, on which no external agent can act, and of which multiple instances cannot be constructed?

At this point, if we were investigating Einstein's criterion in classical physics, we should have to engage with some awkward philosophical issues about the agent's free will. But again, that is not necessary in quantum physics because it is not necessary for the agent to choose θ *freely*: randomly will do, provided that the randomiser is not entangled with $\mathfrak{A}_1$ or $\mathfrak{A}_2$.

So, call the randomiser $\mathfrak{A}_0$. We are now contemplating a *non*-parameterised, isolated network of this form:

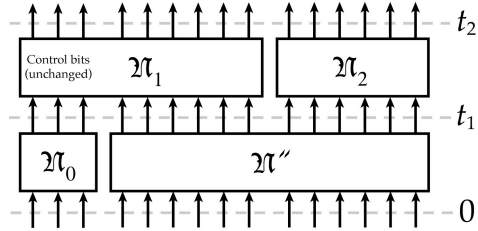

Fig. 3: Using a randomiser as the external agent

where $\mathfrak{A}_0$ transforms $|0\rangle$ to a superposition of eigenstates of the $\hat{z}$-observables (3) of the control qubits, and $\mathfrak{A}_1$ effects a controlled unitary $\sum_i |i\rangle\langle i| \otimes \mathsf{U}_i$, with the control qubits selecting which unitary transformation $\mathsf{U}_i$ will be performed on the remaining qubits of $\mathfrak{A}_1$. The (operator) values of the $\hat{z}$ observables of the control qubits remain unchanged. This loses some generality compared with allowing $\mathfrak{A}''$ to be an arbitrary network acting on all the qubits as in Fig. 1, but the assumption that there are unentangled systems in the past is inherent in the concept of a measurement (see Section 10). The attribute referred to by Einstein as 'independent of what is done with' now means 'the same in all relative states corresponding to different values of the control observables'. Those relative states are all eigenstates of the control observables (at both $t_1$ and $t_2$). And so we continue as in Section 5: we investigate what happens to $\mathfrak{A}_2$ in each relative state with eigenvalue *i*. We find that it is identical to what would happen if the gate acting on $\mathfrak{A}_0$ during $0 < t < t_1$ were replaced by one that transforms $|0\rangle$ to $|i\rangle$ instead of to a superposition of all the eigenstates. Thus we reach the same conclusion as in Section 5.

Now consider Wallace and Timpson's argument in such a case. It hinges on their (mistaken) claim that the only testable predictions made by quantum theory are the expectation values of its observables in one particular state. But in the case of a closed quantum system that has no outside, nothing can detect even the expectation values of most of its observables and therefore, by Wallace and Timpson's argument, those observables do not exist.

So it is Wallace and Timpson's argument, not its rebuttal, that depends on the existence of an external agent.

**9. Algebra-valued reality**

It follows from our proof that according to quantum theory, we are all made of algebra-stuff[4]: the elements of local reality are faithfully described not by real variables or stochastic real variables but by the elements (inappropriately called 'observables') of a certain algebra that can be represented by Hermitian matrices. How it comes about that this algebra-stuff often resolves itself into approximately autonomous channels of information flow ('histories' or 'universes'), most of which approximately obey stochastic classical laws of motion that associate real-number values with the observables, is beyond the scope of this paper (but see Deutsch (2002), Wallace (2011)). For present purposes what matters is that the algebra-stuff often does *not* behave like that, even approximately, and that it especially does not do so in the phenomena of quantum interference and entanglement. Under those circumstances, since there are no real-number values of observables participating in the phenomenon, the *probabilities* of observables taking such values do not exist. If one tries to insist that they do, by interpreting all expectation values in terms of such probabilities, the resulting numbers do not obey the probability calculus (see Deutsch et al. 2000). A failure to understand this can result in various misconceptions about quantum physics in general, and about its locality in particular.

---

[4] Paraphrase of 'the Earth and every living thing are made of star stuff' (Sagan 1980).

Historically the most important of these misconceptions has been that Bell's theorem implies that entangled quantum systems violate Einstein's criterion. As discussed in Deutsch & Hayden (2000), Bell's theorem is about correlations (joint probabilities) of stochastic real variables and therefore does not apply to quantum theory, which neither describes stochastic motion nor uses real-valued observables.

The same error was made by Kastner (2011), who criticised the Deutsch–Hayden proof on the grounds that in separated entangled systems, Bell-inequality-violating correlations 'exist' before the systems are brought together to be compared. But Kastner's 'correlations' are not correlations (probabilities that the real-number values of separate observables are equal), because in the given entanglement phenomena, the information flow is made of algebra-stuff that is not resolving itself into sub-flows each of which is approximately described by real-number variables.

**10. Initial conditions and the arrow of time**

The Deutsch–Hayden proof assumes that when the external agent sets the parameters θ, this has no advanced (backwards-in-time) effect. In terms of the network model, as illustrated in Fig. 1, that means it has no effect on the factual situation of the qubits at $t=0$, nor on the gates and qubits of the sub-network $\mathfrak{N}''$, nor on the timeless Heisenberg state. This causality condition is the usual one in measurement theory. A convenient way of modelling it is to require there to be no entanglement between different qubits at $t=0$. That is the content of taking the initial state to be an eigenstate of the $\hat{z}_a(t)$. Then all entanglement that subsequently exists must be created by interactions between qubits.

It is also possible to model a world in which there *was* some entanglement at $t = 0$ (at the Big Bang, say). We might be reluctant to model this using a prepended network $\mathfrak{A}''$ as in Section 1, in which case we could use an entangled Heisenberg state instead of $|0\rangle$, and again, our proof would go through unchanged. Timpson (2005) argues that in that case, the state would be an element of reality, and a non-local one: 'With the global state of the system still playing such an important role … it is not clear that we have yet gained much in the way of locality'[5]. It is true that such states – including $|0\rangle$ itself – like the laws of motion, the axioms of quantum mechanics, and the global topology of spacetime, are all elements of reality that do not have locations, and are 'non-local' in that sense – a sense in which all theories in physics are 'non-local'. But they also have in common that they cannot be altered. Being universal constants, they are 'independent of what is done to $\mathfrak{S}_1$', and therefore their existence does not violate Einstein's criterion of locality. To avoid confusion, they should be called 'global', not 'non-local'.

For almost any network of gates, some initial states of the qubits would result in evolutions which, if interpreted as modelling physical reality as a whole, would violate the second law of thermodynamics (with suitable coarse-graining and regardless, of course, of which quantum-theoretic picture one uses to describe them). Most would not have a consistent measurement arrow of time. Most would not include the initially-unentangled systems referred to in the operational meanings of observables. Einstein's criterion is invariant under an overall reversal of the measurement arrow of time (since the time-reverse of a quantum network is just another quantum network, and one could always append a network $\mathfrak{A}'''$ to make the final state $|0\rangle$), but what about more exotic, local changes? For example, suppose that in the network $\mathfrak{N}$ of Fig. 2, the measurement arrow of time were locally reversed for $\mathbb{Q}_2$ but not for $\mathbb{Q}_1$. That is to say, the past of $\mathbb{Q}_2$ is affected by 'what is done' to it in the future, rather than vice-versa. Then removing the *not* gate though which $\mathbb{Q}_1$ passes during $0 < t < 1$ would change the observables of $\mathbb{Q}_2$ at $t = 0$. Yet even then, if one adopts the corresponding definition of 'for as long as they remain mutually isolated' (not using cosmological time but following the worldlines of qubits in the direction of the local arrow of time), Einstein's criterion is still satisfied.

---

[5] Timpson makes this comment about what he calls the 'conservative' interpretation of the formalism, as distinct from the 'ontological' interpretation which he considers to be a new theory purporting to replace quantum theory, and of which he has other criticisms. But the distinction between those interpretations, which depends on the alleged difference in measurability between expectation values and operator values of observables, disappears in the light of the comments in Section 5 above.

**11. Conclusions**

Wallace and Timpson's conclusion that there are 'non-local states' in quantum physics is false. Among other things, their critique depends on the false assumption that the only testable predictions made by quantum theory are of expectation values of observables in a particular state.

Einstein's criterion of locality refers to physical systems, not formalisms, and is satisfied by quantum physics regardless of whether it is described in the Schrödinger or Heisenberg picture. Our method of proof is just a method of proof, not a new quantum-theoretic picture nor a variant of quantum theory. Our conclusion that the factual situation of quantum systems is described by their Heisenberg-picture observables is an implication, not an assumption, of the proof.

**Acknowledgments**

I wish to thank David Wallace and an anonymous referee for suggesting improvements to an earlier version of this paper.

**References**

Bekenstein, J.D. 1981 *Universal upper bound on the entropy-to-energy ratio for bounded systems*, Phys. Rev. D23, 287–298

Deutsch, D. and Hayden, P. 2000 *Information flow in entangled quantum systems,* Proc. R. Soc. Lond. **A456** 1759-1774.

Deutsch, D., Ekert, A. & Lupacchini, R. 2000 *Machines, Logic and Quantum Physics*, Bull. Symb. Logic **3** 3.

Deutsch, D. 2002 *The Structure of the Multiverse.* Proc. R. Soc. Lond. **A458** 2911-2923.

Deutsch, D. 2011 *The Beginning of Infinity.* Allen Lane, The Penguin Press, London.

Einstein, A. 1949 quoted in *Albert Einstein: Philosopher, Scientist,* P.A. Schilpp, Ed., Library of Living Philosophers, Evanston, 3rd edition (1970), p85.

Everett, H. 1957 *On the Foundations of Quantum Mechanics*, Ph.D. thesis, Princeton University.

Kastner, R.E. 2011 *Quantum Nonlocality: Not Eliminated by the Heisenberg Picture* Found. Phys. **41** 7, 1137-1142.

Rubin, M.A. 2002, *Locality in the Everett Interpretation of Quantum Field Theory* Found. Phys. **32** 1495-523.

Rubin, M.A. 2011 *Observers and Locality in Everett Quantum Field Theory* Found. Phys. **41** 1236-62.

Sagan, C. 1980 *Cosmos: A Personal Voyage*, Public Broadcasting Service.

Timpson, C.J. 2005 *Nonlocality and information flow: The approach of Deutsch and Hayden.* Found. Phys. **35** 2 313-43

Wallace, D. 2011 *The Everett Interpretation.* Oxford University Press.

Wallace, D. and Timpson, C.J. 2007 *Non-locality and gauge freedom in Deutsch and Hayden's formulation of quantum mechanics.* Found. Phys. Lett. **37** 951-955.